\documentclass{PoS}
\usepackage{graphicx}     
\usepackage[dvips]{epsfig}
\newcommand{\bea}{\begin{eqnarray}}   
\newcommand{\eea}{\end{eqnarray}}

\newcommand{\tetatt}{\mbox{$\theta_{23}$}}
\newcommand{\deltaunodue}{\mbox{$\Delta m_{12}^2 $}}

\newcommand{\deltaduetre}{\mbox{$\Delta m_{23}^2 $}}
\newcommand{\tetaot}{\mbox{$\theta_{13}$}}

\newcommand{\beq}{\begin{eqnarray}}
\newcommand{\eeq}{\end{eqnarray}}
\newcommand{\numu}{\mbox{$\nu_{\mu}$}}
\newcommand{\numubar}{\mbox{$\overline{\nu}_{\mu}$}}
\newcommand{\nue}{\mbox{$\nu_{e}$}}
\newcommand{\nuebar}{\mbox{$\overline{\nu}_{e}$}}

\newcommand{\delot}{\mbox{$\Delta_{23}$}}

\font\tenrsfs=rsfs10 at 12pt
\font\sevenrsfs=rsfs7
\font\fiversfs=rsfs5
\newfam\rsfsfam
\textfont\rsfsfam=\tenrsfs
\scriptfont\rsfsfam=\sevenrsfs
\scriptscriptfont\rsfsfam=\fiversfs
\def\mathscr#1{{\fam\rsfsfam\relax#1}}

\makeatletter
\def\art{\@ifnextchar[{\eart}{\oart}}
\def\eart[#1]#2#3#4#5#6{{\rm #2}, {#3 #4} {\rm (#6) #5} [{{#1}}]}
\def\hepart[#1]#2{{\rm #2, {#1}}}
\newcommand{\oart}[5]{{\rm #1}, {#2 #3} {\rm (#5) #4}}

\def\circa#1{\,\raise.3ex\hbox{$#1$\kern-.75em\lower1ex\hbox{$\sim$}}\,}

\title{Optimisation of future long baseline neutrino experiments}
\ShortTitle{Optimisation of future long baseline neutrino experiments}

\author{Olga Mena \thanks{It is a pleasure to thank all my long-baseline collaborators and to Enrique Fern\'andez Mart\'{\i}nez who has produced the sensitivity plots shown in this talk.
O.~M is supported by a \emph{Ram\'on y Cajal} contract from MEC, Spain.
}\\
        IEEC/CSIC, Bellaterra, Spain and\\ 
	IFIC/CSIC, Valencia, Spain\\
        E-mail: \email{mena@ieec.uab.es}}

\abstract{
The aim of this \emph{talk} is to review near and far future long baseline neutrino experiments as superbeams, $\beta$-Beams and neutrino factories, comparing 
their sensitivities to the unknown parameters in the neutrino oscillation sector. We focus on the extraction of the neutrino mass hierarchy, exploring alternatives to the commonly used neutrino-antineutrino comparison. Special attention to a new concept of neutrino factory design, the \emph{low energy neutrino factory}, is given.}

\FullConference{10th International Workshop on Neutrino Factories,
                 Super beams and Beta beams\\
		 June 30- July 5, 2008\\
		 Valencia, Spain\\
		 --\\}
\begin{document}

\section{Introduction}
During the last several years the physics of neutrinos has achieved
remarkable progress. The present data require two large
($\theta_{12}$ and $\theta_{23}$) and one small ($\theta_{13}$) angles
in the neutrino mixing matrix, and at least two mass squared differences,
$\Delta m_{ij}^{2} \equiv m_j^2 -m_i^2$ (where $m_{j}$'s are the neutrino
masses), one driving the atmospheric ($\deltaduetre$) and the other one the solar ($\deltaunodue$) neutrino oscillations. The mixing
angles $\theta_{12}$ and $\theta_{23}$ control the solar and the
atmospheric neutrino oscillations, while $\theta_{13}$ is the angle
which connects the atmospheric and solar neutrino realms.

A recent global fit~\cite{concha} provides the following $3 \sigma$ allowed ranges for the atmospheric mixing parameters $|\deltaduetre| =(2 - 3.2)\times10^{-3}$~eV$^{2}$ and $0.32<\sin^2\theta_{23}<0.64$. The sign of $\deltaduetre$, sign$(\deltaduetre)$, cannot be determined with the existing data. The two possibilities, $\deltaduetre > 0$ or $\deltaduetre < 0$, correspond to two different types of neutrino mass ordering: normal hierarchy and inverted hierarchy.
In addition, information on the octant in which $\theta_{23}$ lies, if $\sin^22\theta_{23} \neq 1$, is beyond the reach of present experiments. The solar neutrino oscillation parameters lie in the low-LMA (Large Mixing Angle) region, with best fit values~\cite{concha} $\deltaunodue =7.9 \times 10^{-5}~{\rm eV^2}$ and $\sin^2 \theta_{12} =0.30$. A combined 3-neutrino oscillation analysis of the solar, atmospheric, reactor and long-baseline neutrino data~\cite{concha} constrains the third mixing angle to be $\sin^2\theta_{13} < 0.04$ at the $3\sigma$ C.L. 

The future goals for the study of neutrino properties is to precisely determine the already measured oscillation parameters
and to obtain information on the unknown ones: namely $\theta_{13}$,
the CP--violating phase $\delta$ and the type of neutrino mass
hierarchy (or equivalently sign$(\deltaduetre)$). 
\section{The golden channels}

The most promising way to determine the unknown parameters $\delta$ and $\tetaot$ is through the detection of the subleading transitions $\nue (\nuebar) \leftrightarrow \numu (\numubar)$. These channels have been named as \emph{golden channels}~\cite{golden}. Defining $\Delta_{ij} \equiv \frac{\Delta m^2_{ij}}{2 E}$, a convenient and precise approximation is obtained by expanding to second order in the following small parameters: 
 $\tetaot$, $\Delta_{12}/\Delta_{23}$, $\Delta_{12}/A$ and $\Delta_{12} \, L$~\cite{golden,ahk}
\bea
P_{\nu_ e \nu_\mu ( \bar \nu_e \bar \nu_\mu ) } & = & 
s_{23}^2 \sin^2 2 \tetaot \, \left ( \frac{ \delot }{ \tilde B_\mp } \right )^2
   \, \sin^2 \left( \frac{ \tilde B_\mp \, L}{2} \right) \, + \, 
c_{23}^2 \sin^2 2 \theta_{12} \, \left( \frac{ \Delta_{12} }{A} \right )^2 
   \, \sin^2 \left( \frac{A \ L}{2} \right ) \nonumber \\
& + & \label{approxprob}
\tilde J \; \frac{ \Delta_{12} }{A} \, \frac{ \delot }{ \tilde B_\mp } 
   \, \sin \left( \frac{ A L}{2}\right) 
   \, \sin \left( \frac{\tilde B_{\mp} L}{2}\right) 
   \, \cos \left( \pm \delta - \frac{ \delot \, L}{2} \right ) \, , 
\label{eqn:hastaelmogno}
\eea
where $L$ is the baseline, $\tilde B_\mp \equiv |A \mp \delot|$ and the 
matter parameter $A$ is defined in terms of the average electron 
number density, $n_e(L)$,  as $A \equiv \sqrt{2} \, G_F \, n_e(L)$, and $\tilde J$ is defined as 
\beq
 \tilde J \equiv \cos \theta_{13} \; \sin 2 \theta_{13}\; \sin 2 \theta_{23}\;
 \sin 2 \theta_{12}~.
\eeq
The golden transitions Eqs.~(\ref{eqn:hastaelmogno}) are sensitive to all the unknown parameters quoted in the introductory section. They are clearly sensitive to the mixing angle $\theta_{13}$. These channels are also sensitive to the CP--violating phase (via the third term or the \emph{interference} term, the only term which differs for neutrinos and antineutrinos). The golden transitions allow us also to extract the sign of the atmospheric mass difference exploiting matter effects. In the presence of matter effects, the neutrino (antineutrino) oscillation probability gets enhanced~\cite{matter} for the normal (inverted) hierarchy. Making use of the different matter effects for neutrinos and antineutrinos seems, in principle, the most promising way to distinguish among the two possibilities: normal versus inverted hierarchy.

\section{Degenerate solutions}
We can ask ourselves whether it is possible to unambiguously determine $\delta$ and $\tetaot$ by measuring the golden transition probabilities, $\nue \to \numu$ and $\nuebar \to \numubar$, Eqs.~(\ref{eqn:hastaelmogno}) (or equivalently,  $\numu \to \nue$ and $\numubar \to \nuebar$) at fixed neutrino energy $E$ and at just one baseline $L$. The answer is no. By exploring the full (allowed) range of the $\delta$ and $\tetaot$ parameters, that is, $-180^{\circ}<\delta<180^{\circ}$ and $0^{\circ}<\tetaot<10^{\circ}$, one finds, at fixed neutrino energy and at fixed baseline, the existence of degenerate solutions ($\theta^{'}_{13}$, $\delta^{'}$), that we label \emph{intrinsic degeneracies}, which give the same oscillation probabilities than the set ($\tetaot$, $\delta$) chosen by nature~\cite{burguet}. It has also been pointed out that other fake solutions might appear from  unresolved degeneracies in two other oscillation parameters:
\begin{enumerate}
\item The sign of the atmospheric mass difference $\Delta m_{23}^2$ may remain unknown. In this particular case, $P (\theta^{'}_{13}, \delta^{'}, -\Delta m_{23}^{2}) = P (\theta_{13}, \delta, \Delta m_{23}^2)$~\cite{sign1,sign2}.
\item Disappearance experiments only give us information on $\sin^{2} 2 \theta_{23}$: is $\theta_{23}$ in the first octant, or is it in the second one, $(\pi/2 -\theta_{23}$)? . In terms of the probabilities, $P (\theta^{'}_{13}, \delta^{'}, \frac{\pi}{2}-\tetatt) = P (\theta_{13}, \delta ,\tetatt)$~\cite{sign2,th231}.
\end{enumerate}
All these ambiguities complicate the experimental determination of $\delta$ and
$\tetaot$. The situation has been dubbed the \emph{eight-fold degeneracy}. A lot of work has been devoted to resolve the degeneracies by exploiting the different neutrino energy and baseline dependence of two (or more) LBL experiments. A complete list of references is beyond the scope of this talk. I suggest to see Ref.~\cite{iss} and references therein.

\section{The tree level approach: superbeams}

The next generation of long baseline $\nu_e$ neutrino appearance experiments will be the so-called superbeam experiments. The major goal of superbeam experiments is to set a non-zero value of the small mixing angle $\theta_{13}$ (or, in the absence of a positive signal, to improve the current upper bound
on this mixing angle).
A superbeam experiment consists, basically, of a higher intensity version of a conventional neutrino (antineutrino) beam. Superbeams represent the logical next step in accelerator-based neutrino physics.  There are two possible strategies regarding the neutrino beam. The \emph{off-axis} technique produces a neutrino spectrum very narrow in energy (nearly monochromatic, $\Delta E /E \sim 15 - 25\%$), which peaks at lower energies with respect to the on-axis one. The off-axis technique allows a discrimination between the peaked $\nu_e$ oscillation signal and the intrinsic $\nu_e$ background which has a broad energy spectrum. In addition, the off-axis technique reduces significantly the background resulting from neutral current interactions of higher energy neutrinos with a $\pi^{0}$ in the final state. Unfortunately, off axis experiments are counting experiments in which one has only two measurements (the number of neutrinos and the number of antineutrino events) and resolving degeneracies becomes an impossible task. This is the technique exploited by the $\nu_e$ appearance experiments T2K~\cite{T2K} and NO$\nu$A~\cite{newNOvA}. 

The \emph{wide band beam (WBB)} technique exploits the spectral information of the signal, being sensitive to many $E/L$'s at the same time. The neutrino beam is on-axis and therefore the fluxes and the beam energies are higher than the ones exploited in the off-axis case. Higher beam energies imply longer distances, and therefore larger detectors. The WBB technique requires Mton class detectors with extremely good energy resolution and optimal neutral-current background rejection. See Refs.~\cite{dusel} for the physics opportunities with a WBB at a Deep Underground Science and Engineering Laboratory (DUSEL). 

The authors of \cite{comparison} have studied carefully the two possible techniques, finding (for the same exposure) the WBB option better for the mass hierarchy extraction, and the NO$\nu$A off-axis experiment better for CP--violation searches.

\section{The race for the hierarchy}
\label{sec:strat}
\begin{figure}[t]
\begin{center}
\begin{tabular}{lll}
\includegraphics[width=2in]{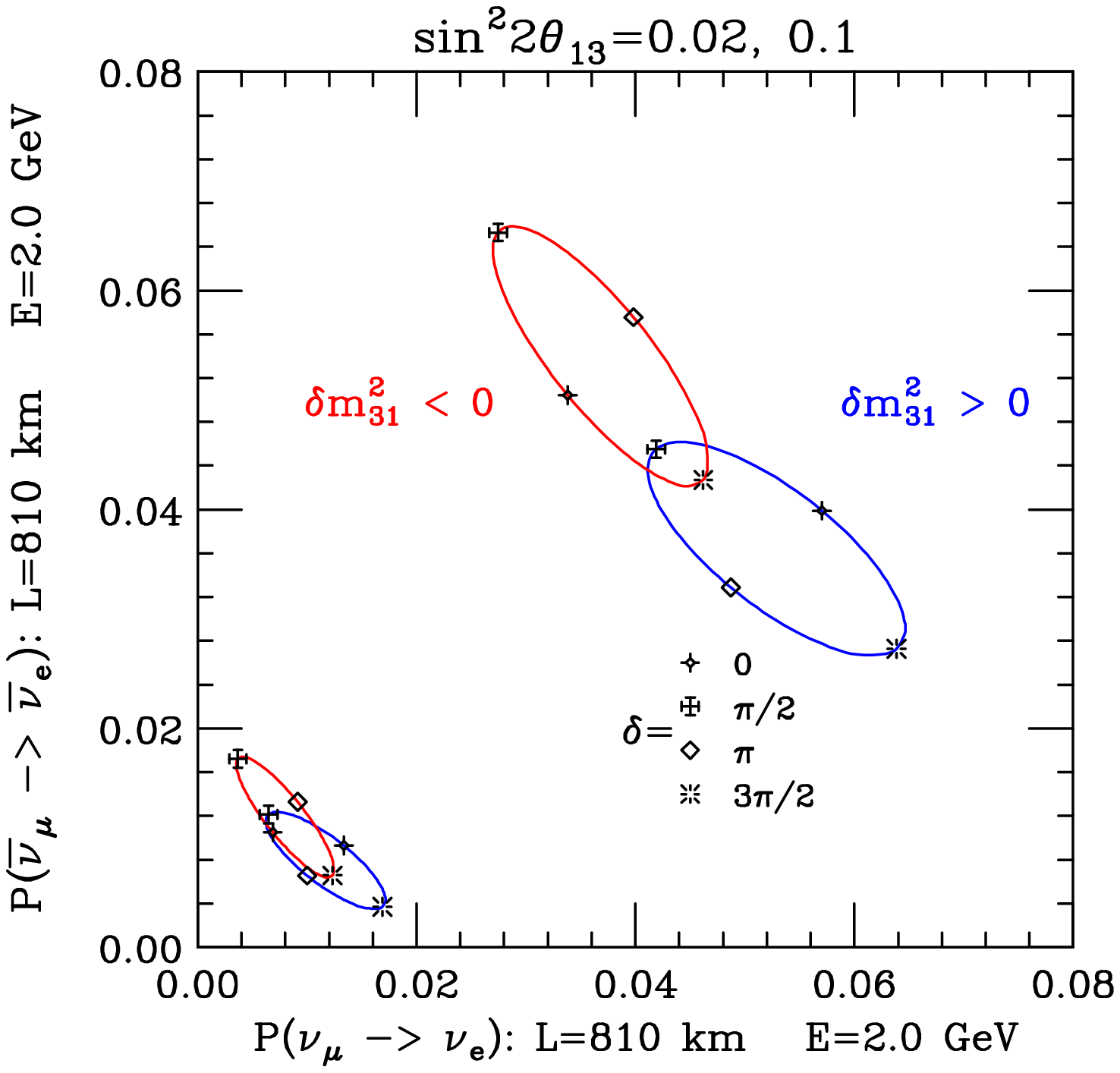}&\hskip 0.cm
\includegraphics[width=2in]{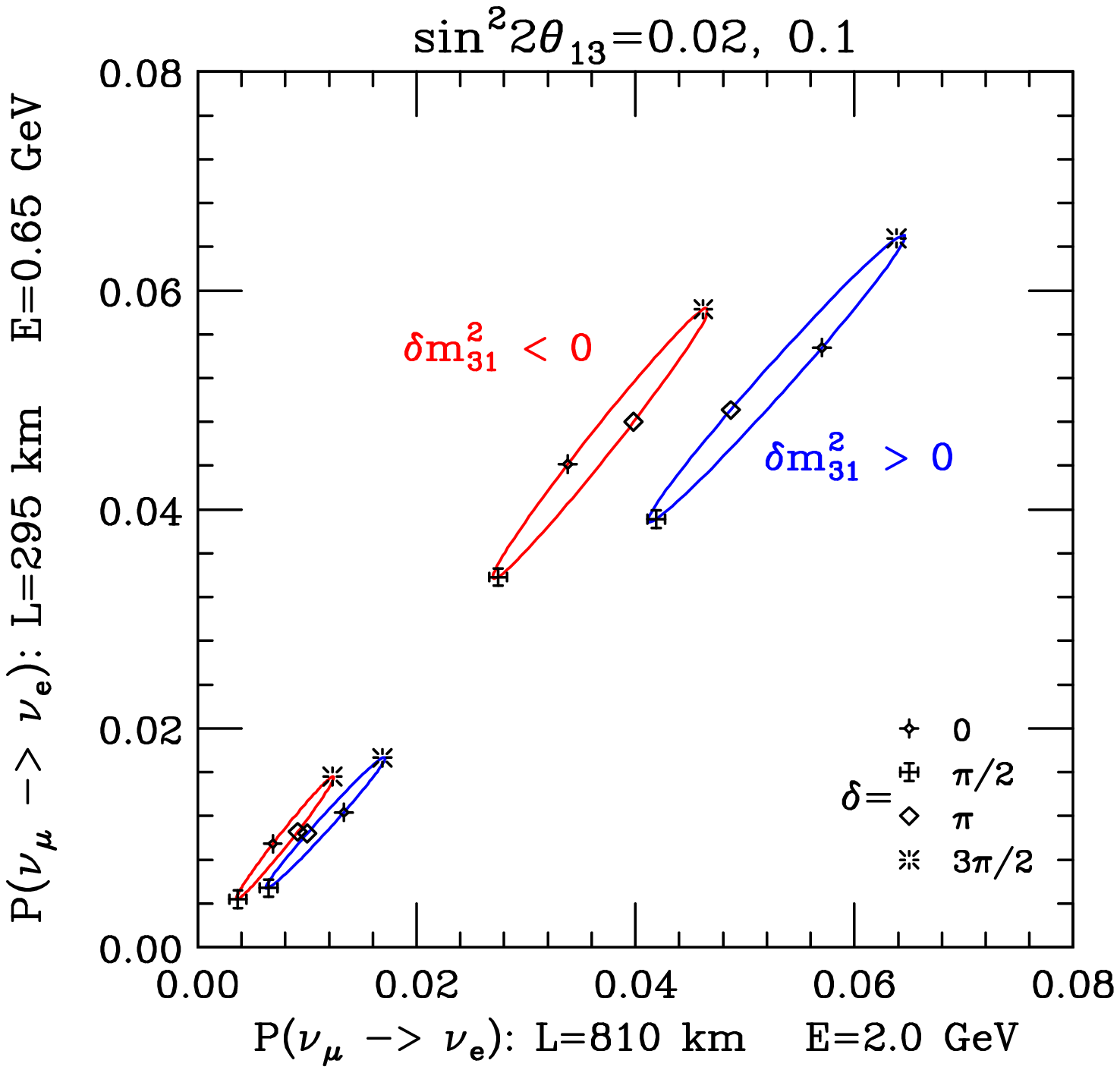}&\hskip 0.cm
\includegraphics[width=2in]{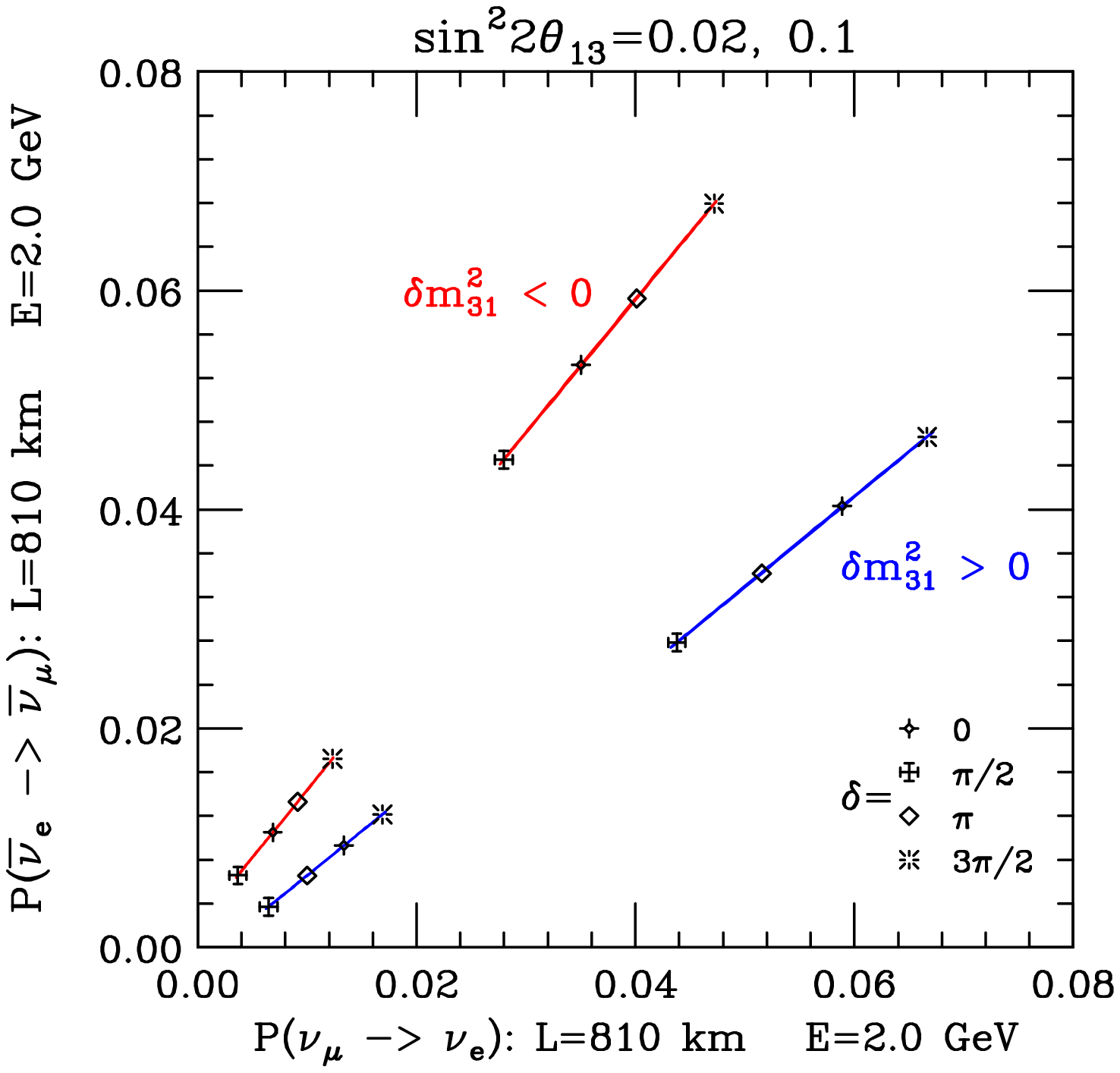}\\
\end{tabular}
\end{center}
\caption[]{\textit{Left panel: bi--probability plot for $P(\nu_\mu \to \nu_e)$ versus $P(\bar{\nu}_\mu \to \bar{\nu}_e)$ at a baseline of 810 km and an energy of 2.0~GeV for the normal (blue) and the inverted (red) hierarchies. The smaller, lower (larger, upper) ellipses are for $\sin^2 2 \theta_{13}=0.02$ ($~0.10$).
Medium panel: bi--probability plot for $P(\nu_\mu \to \nu_e)$ versus $P(\nu_\mu \to \nu_e)$ with baselines 295 km and 810 km. The mean neutrino energies are chosen such that the $\langle E \rangle /L$ for the two experiments are approximately identical. The right panel is the bi--probability plot for $P(\nu_\mu \to \nu_e)$ versus $P(\bar{\nu}_e \to \bar{\nu}_\mu)$ for the normal (blue) and the inverted (red) hierarchies. The baseline and mean neutrino energy for both experiments are 810 km and  $\sim$ 2~GeV.} }
\label{fig:comp2}
\end{figure}

Typically, the proposed near term LBL neutrino oscillation experiments (superbeams) have a single far detector and plan to run with the beam in two different modes, muon neutrinos and muon antineutrinos. Suppose we compute the oscillation probabilities $P(\nu_ \mu \to \nu_e)$ and $P(\bar \nu_\mu \to \bar \nu_e)$ for a given set of oscillation parameters and the CP--phase $\delta$ is varied between $0$ and $2 \pi$: we obtain a closed CP trajectory (an ellipse) in the bi--probability space of neutrino and antineutrino conversion~\cite{MN01}. Matter effects are responsible for the departure of the center of the ellipses from the diagonal line in the bi--probability plane for normal and inverted hierarchy. In Fig.~\ref{fig:comp2}, we have illustrated the case for $E=2.0$~GeV and $L=810$~km, which roughly correspond to those of the NO$\nu$A experiment. The distance between the center of the ellipse for the normal hierarchy (lower blue) and that for  the inverted hierarchy (upper red) is governed by the size of the matter effects. Notice that the ellipses overlap for a significant fraction of values of the CP--phase $\delta$ for every allowed value of $\sin^2 2
\theta_{13}$.  This makes the determination of sign$(\deltaduetre)$ extremely difficult, i.~e., the sign$(\deltaduetre)$-extraction is not free of degeneracies and it is highly dependent on the value of $\delta$. 

Following the line of thought developed by Minakata, Nunokawa and Parke~\cite{MNP03}, we exploited~\cite{mmnp} the neutrino data only from two experiments at different distances and at different off-axis locations, such that the $\langle E \rangle /L$ is the same for the two experiments (see also Refs.~\cite{HLW02,BMW02, SN1,twodetect}). In the case of bi--probability plots for neutrino--neutrino modes at different distances (which will be referred as near (N) and far (F)), the CP--trajectory is also elliptical. 
In Fig.~\ref{fig:comp2} (medium panel) we present the bi--probability plot for the mean energies and baselines of the $\nu_e$ appearance experiments T2K and NO$\nu$A. The overlap of the two ellipses, which implies the presence of a degeneracy of the type of hierarchy with other parameters, is determined by their width and the difference in the slopes. The ratio of the slopes, at first order in the matter parameter, and assuming that the $\langle E \rangle /L$ of the near and far experiments is the same, reads
\beq
\frac{\alpha_+}{\alpha_-} \simeq
1 +  2 \left( A_{\rm N} L_{\rm N} - A_{\rm F} L_{\rm F} \right)\left( \frac{1}{\Delta_{13} L/2} - \frac{1}{\tan(\Delta_{13}L/2)} \right)~,
\label{eq:ratioapp}
\eeq
where $\alpha_+$ and $\alpha_-$ are the slopes of the center of the ellipses as one varies $\theta_{13}$ for normal and inverted
hierarchies, $A_{\rm F}$ and $A_{\rm N}$ are the matter parameters, and $L_{\rm F}$ and $L_{\rm N}$ are the baselines for the two experiments. 
The separation between the center of the ellipses for the two hierarchies increases as the difference in the matter parameter times the path length, ($AL$), 
for the two experiments increases. 
However the width of the ellipses is crucial: even when the separation 
between the central axes of the two regions is substantial, if the ellipses  for
the normal and inverted hierarchy overlap, the hierarchy cannot be resolved for
values of the CP--phase, $\delta$, for which there is overlap. The width of the ellipses is determined by the difference in the $\langle E \rangle /L$ of the two experiments.

In the case of bi--probability plots for the $\nu_\mu \to \nu_e$ and its CPT conjugated channel $\bar{\nu}_e \to \bar{\nu}_\mu$ at the same energy divided by baseline,$\langle E \rangle /L$, the CPT--trajectory collapses to a line (see Fig.~\ref{fig:comp2}, right panel). 
As for the neutrino-neutrino case, assuming that the $\langle E \rangle /L$ of the CPT conjugated channels is the same (to minimize the ellipses width), at first order, the ratio of the slopes reads~\cite{MNP03}
\beq
\frac{\alpha_+}{\alpha_-} \simeq
1 +  2 \left( A L + A_{\rm CPT} L_{\rm CPT} \right)\left( \frac{1}{\Delta_{13}L/2} - \frac{1}{\tan(\Delta_{13}L/2)} \right)~,
\label{eq:ratiocpt}
\eeq
where $\alpha_+$ and $\alpha_-$ are the slopes of the center of the ellipses as one varies $\theta_{13}$ for normal and 
inverted hierarchies, $A$ and $A_{\rm CPT}$ are the matter parameters and $L$ and $L_{\rm CPT}$ are the baselines for the
 two experiments which exploit the $\nu_\mu \to \nu_e$ and its CPT conjugated channel ($\bar{\nu}_e \to \bar{\nu}_\mu$). Notice that, compared to the neutrino--neutrino case given by Eq.~(\ref{eq:ratioapp}), the separation between the center of the ellipses for the two hierarchies increases as the
 sum of the matter parameter times the baseline, $AL$, for both experiments does. Here the ratio of the slopes is enhanced by matter effects for both $\nu_\mu \to \nu_e$ and its CPT conjugated channel $\bar{\nu}_e \to \bar{\nu}_\mu$. Figure~\ref{fig:comp2} (right panel) shows the bi--probability curves for the combination of these two flavor transitions, assuming that the two experiments are performed at the same mean energy and baseline. If the $\langle E \rangle /L$ of both experiments is the same, the ellipses will become lines with a negligible width. The separation of the lines for the normal and inverted hierarchy grows as the matter effects for both experiments increase.    
Consequently, the comparison of CPT conjugated channels is more sensitive to the neutrino mass hierarchy than the neutrino--neutrino one, see Ref.~\cite{cptus}.

\section{Higher order corrections: $\beta$-beams and neutrino factories}
\label{sec:beta}
Precision lepton flavor physics requires powerful machines and extremely pure neutrino beams. Future LBL experiments which exploits pure $\nu_e$ ($\bar{\nu}_e$) neutrino beams are $\beta$-beams and neutrino factories.
A $\beta$-beam experiment~\cite{zucchelli} exploit ions which are accelerated to high Lorentz factors, stored and then $\beta$-decay, producing a collimated electron neutrino beam. The typical neutrino energies are in the $200$~MeV-GeV range, requiring detectors with hundred-of-MeV thresholds and good energy resolution. The only requirement is good muon identification in order to detect the appearance of muon neutrinos (or muon antineutrinos) from the initial electron neutrino (or antineutrino) beam. No magnetisation is required and therefore several detectors technologies (water Cherenkov, totally active scintillator (TASD), liquid argon and non-magnetised iron calorimeter) could be used, depending on the peak energy. 

The initial $\beta$-beam setup~\cite{zucchelli} considers a \emph{low}-$\gamma$ machine which accelerates $^6He$ ($\bar{\nu}_e$ emitter) and $^{18}Ne$ ($\nu_e$ emitter) up to $\gamma \sim 100$. In order to tune the $E/L$ at the vacuum oscillation maximum, a large water Cherenkov detector is located at a distance $\mathcal{O}(100)$~km. The first exciting option to improve this initial $\beta$-beam scenario was presented in Ref.~\cite{betabeampilar}, where the possibility of using higher $\gamma$ factors was first suggested. The second exciting option, see Ref.~\cite{rubbia}, proposes to accelerate alternative ions, as $^8Li$ ($\bar{\nu}_e$ emitter) and $^{8}B$ ($\nu_e$ emitter), with higher Q-values. A plethora of setups have been proposed in the literature (see Ref.~\cite{iss} for a complete list of references).

A neutrino factory (NF)~\cite{geer,dgh} consists, essentially, of a muon storage ring with long straight sections along which the muons decay. These muons provide high intensity and extremely pure neutrino beams. Hence, the NF provides $\nue$ and $\nuebar$ beams in addition to $\numu$ and $\numubar$ beams, with minimal systematic uncertainties on the neutrino flux and spectrum. One of the most important advantages of the NF, compared to the $\beta$-beam, is its ability to measure with high precision the atmospheric mixing parameters via the disappearance channels ($\numu (\numubar)\to \numu (\numubar)$)

The NF exploits the golden signature of the \emph{wrong-sign muon} events~\cite{golden}. What is a ``wrong sign muon'' event? Suppose, for example, that positive charged muons have been stored in the ring. These muons will decay as $\mu^{+}\to e^{+} + \nue + \numubar$. The muon antineutrinos will interact in the detector to produce positive muons. Then, any \emph{wrong-sign muons}
 (negatively-charged muons) detected are an unambiguous proof of electron neutrino oscillations in the $\nue\to\numu$ channel. A magnetized detector with good muon charge identification is mandatory. 

In Ref.~\cite{iss} a complete study of possible near and far future LBL facilities has been performed, including superbeams, $\beta$-beams and NF. The optimal setup is found to be a $20$~GeV NF delivering $5 \cdot 10^{20}$ muon decays per year, baseline and polarity. The running time assumed is 5 years per polarity. Two iron calorimeter detectors of $50$~kton are placed at two different baselines, at $\mathcal{O}(4000)$~km and at $\mathcal{O}(7000)$~km (the so-called \emph{magic baseline}~\cite{magic}). The oscillated data from detector at the largest baseline helps enormously in resolving the mass hierarchy degeneracy. Once that the sign($\Delta m_{23}^2$) degeneracy is resolved, leptonic CP--violation can be measured unambiguously using the data from the detector located at $\mathcal{O}(4000)$~km. 
\subsection{The low energy neutrino factory}
The optimal $20$~GeV plus two detectors NF setup described in the previous section outperforms any other planned scenario so far, as we will shortly show. However, such an aggressive setup could be extremely challenging to construct (a $\mathcal{O}(7000)$~km baseline would require the construction of a decay tunnel with an inclination of $\sim 30^\circ$~\footnote{I would like to thank C.~Quigg for making this observation}). More important, such an aggressive scenario might not be needed if $\sin^2 2 \theta_{13} > 10^{-4}-10^{-3}$. The reason for that is simple: if the mixing angle $\theta_{13}$ is not so small, there is no need to go to very long baselines to amplify it. Shorter baselines require lower energies. 
Lower energy Neutrino Factories (LENF), which store muons with energies $ < 10$~GeV, require a detector technology that can detect lower energy muons. In previous studies~\cite{lownf1,lownf2}, we have considered a LENF with an energy of about $4$~GeV providing $5 \cdot 10^{20}$ muon decays per year. The detector exploited was a magnetized TASD~\cite{lownf2} of $20$~kton, with a muon energy detection threshold of $500$~MeV, located at a distance of $1480$~km (Fermilab to Henderson mine). The results are similar for a baseline of $1280$~km (Fermilab to Homestake). The intrinsic background fraction is $10^{-3}$.
Here, we improve the LENF setup in two ways. First, the detector energy resolution would be $dE/E \sim 10\%$~\footnote{Based on NO$\nu$A results, we expect the TASD dE/E to be better than $6\%$ at 2~GeV.}.
Second, and more interesting, since it seems possible to measure in a magnetized TASD the electron charge~\cite{alan}, apart of exploiting the $\nue (\nuebar) \to \numu (\numubar)$ channels, their \emph{T}-conjugate channels $\numu (\numubar) \to \nue (\nuebar)$ channels are also added to the analysis. These extra \emph{T}-conjugate channels will help enormously in resolving degeneracies. We assume here that the electron charge identification is constant in energy and equal to $50\%$ (a detailed analysis will be presented elsewhere~\cite{alan}).

\begin{figure}[h]
\begin{center}
\begin{tabular}{lll}
\includegraphics[width=2in]{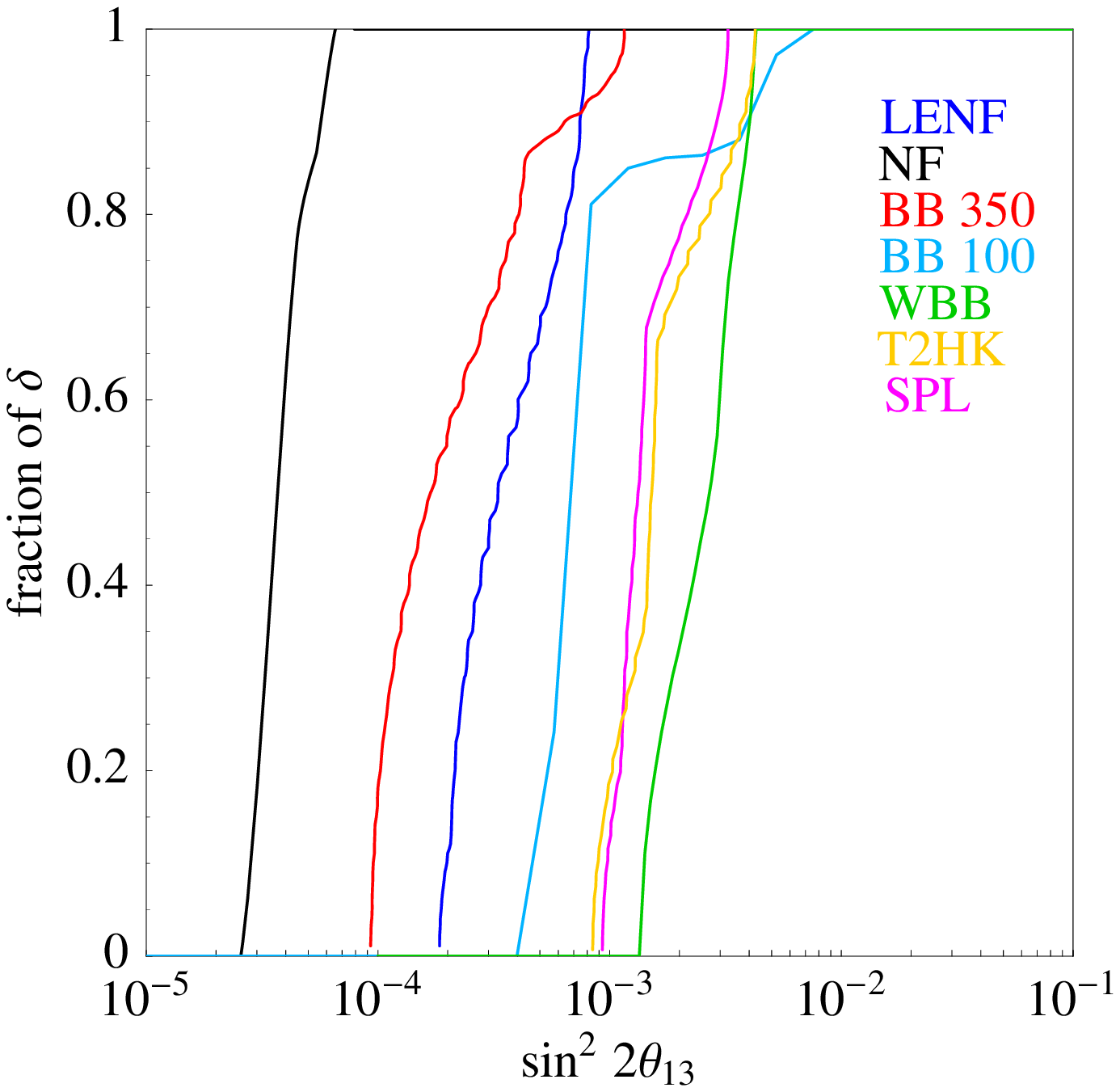}&\hskip 0.cm
\includegraphics[width=2in]{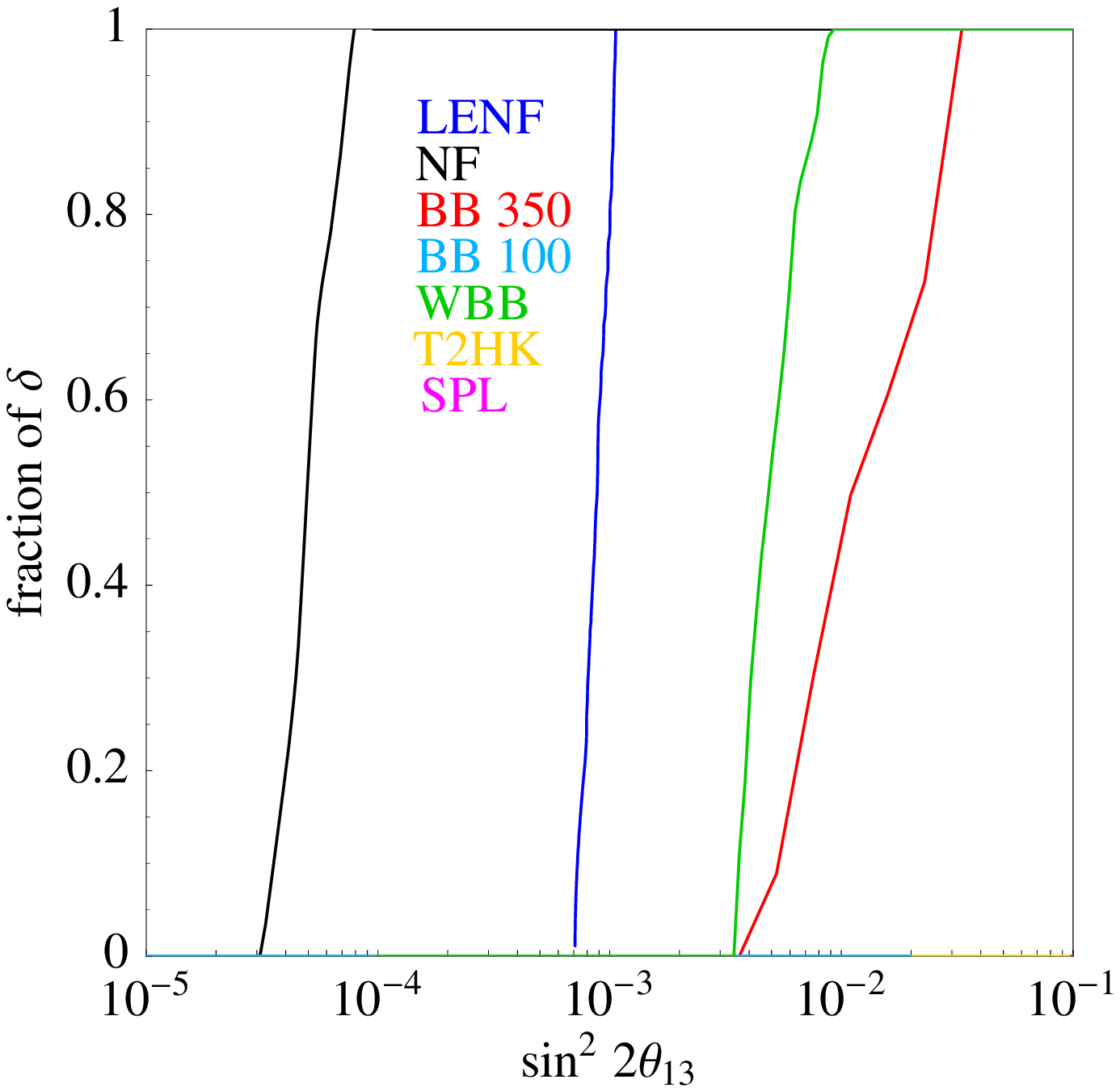}&\hskip 0.cm
\includegraphics[width=2in]{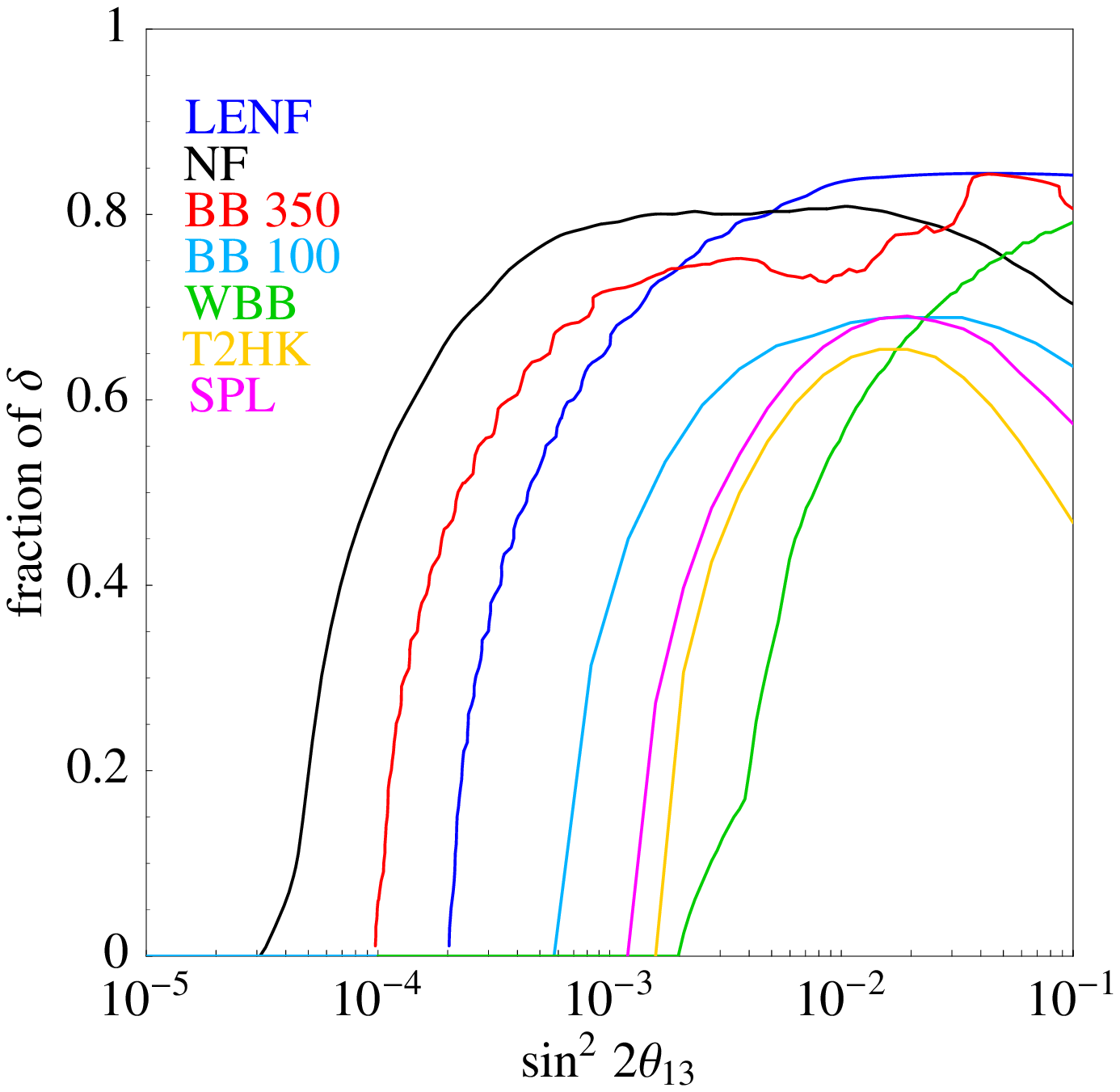}\\
\end{tabular}
\end{center}
\caption{\textit{The left, medium and right panels show the 3-$\sigma$ $\theta_{13}$ discovery potential, the mass hierarchy sensitivity and the CP-violation sensitivity, respectively, expected from future data at a LENF. We present as well the expected sensitivities from future data at the different LBL experiments presented in Ref.~\cite{iss}. Figure produced using the GLOBES software~\cite{globes}.}}
\label{fig:compx}
\end{figure}

Figure~\ref{fig:compx} shows the 3-$\sigma$ $\theta_{13}$ discovery potential and the sensitivities to the mass hierarchy to CP--violation expected from data at  a future LENF with the characteristics quoted above (the exposure is $10^{23}$ kton-decays). As a comparison, we show as well the expected sensitivities exploiting data from the future LBL facilities presented in Ref.~\cite{iss}. Notice that the high-$\gamma$ $\beta$-beam~\cite{betabeampilar}, labelled as BB$350$, provides a slightly better sensitivity than the LENF to both CP violation and to $\theta_{13}$ due to its lower energy and its huge statistics. The $20$~GeV NF with two baselines ($4000$~km+$7000$~km) is unbeatable, but we might only need such an aggressive scenario if $\sin^2 2 \theta_{13} < 10^{-4}-10^{-3}$.
To conclude, the low energy neutrino factory (LENF)~\cite{lownf1,lownf2} provides a compromise between super precision machines and feasible setups, and it could provide an ideal and realistic scenario for precision lepton physics.  


\end{document}